\pdfoutput=1
\RequirePackage{ifpdf}
\ifpdf
\documentclass[pdftex]{sigma}
\else
\documentclass{sigma}
\fi

\def\o#1{\ensuremath \mathbf{#1}}
\newcommand{\Sym}{\ensuremath \mathrm{Sym}}

\newcommand{\rk}{\ensuremath \mathop \mathrm{rank} \nolimits}

\begin{document}
\allowdisplaybreaks

\renewcommand{\thefootnote}{$\star$}

\renewcommand{\PaperNumber}{002}

\FirstPageHeading

\ShortArticleName{Invertible Darboux Transformations}

\ArticleName{Invertible Darboux Transformations\footnote{This
paper is a~contribution to the Special Issue ``Symmetries of Dif\/ferential Equations: Frames,
Invariants and Applications''.
The full collection is available
at
\href{http://www.emis.de/journals/SIGMA/SDE2012.html}{http://www.emis.de/journals/SIGMA/SDE2012.html}}}

\Author{Ekaterina SHEMYAKOVA}

\AuthorNameForHeading{E.~Shemyakova}

\Address{Department of Mathematics, SUNY at New Paltz, 1 Hawk Dr. New Paltz, NY 12561, USA}
\Email{\href{mailto:shemyake@newpaltz.edu}{shemyake@newpaltz.edu}}

\ArticleDates{Received October 01, 2012, in f\/inal form January 01, 2013; Published online January 04, 2013}

\Abstract{For operators of many dif\/ferent kinds it has been proved that (generalized) Darboux
transformations
can be built using so called Wronskian formulae.
Such Darboux transformations are not invertible in the sense
that the corresponding mappings of the ope\-ra\-tor kernels are not invertible.
The only known invertible ones
were Laplace transformations (and their compositions), which are special cases of Darboux
transformations
for hyperbolic bivariate operators of order~2.
In the present paper we f\/ind a~criteria for a~bivariate linear partial dif\/ferential operator of an
arbitrary order $d$
to have an invertible Darboux transformation.
We show that Wronkian formulae may fail in some cases,
and f\/ind suf\/f\/icient conditions for such formulae to work.}


\Classification{37K10; 37K15}

\renewcommand{\thefootnote}{\arabic{footnote}}
\setcounter{footnote}{0}

\section{Introduction}
\label{sec:intr}

Darboux transformations are shape preserving transformations of linear partial dif\/ferential
operators.
These transformations were originally introduced in~\cite{Darboux2} f\/irst for operators
\begin{gather}\label{op:L2XX}
-D_x^2+u-\lambda,
\end{gather}
corresponding to Sturm--Liouville equations, where $u\in K$, and where $K$ is some dif\/ferential
f\/ield (see Section~\ref{sec:pre}),
and $\lambda$ is a~constant, and, secondly, for operators of the form
\begin{gather}\label{op:L2}
D_xD_y+a D_x+bD_y+c,
\end{gather}
where $a,b,c\in K$, which is a~form equivalent up to a~change of variable to
stationary 2D Schr\"odinger operator.
These transformations were suggested as a~possible method for the solution of the corresponding
linear
partial dif\/ferential equations\footnote{Though we cannot guarantee in advance that
it will be successful.}.
By exploiting these ideas many new kinds of integrable
1D and 2D Schr\"odinger equations have been discovered.
See, for example, this inf\/luential paper~\cite{novikov1997exactly}.

Decades later, Darboux transformations became a~standard tool of the inverse scattering method,
where they are applied to the linear partial dif\/ferential operators forming the Lax pair and thus
eventually serve
for the solution of nonlinear partial dif\/ferential
equations~\cite{matveev:1991:darboux,rogers_schief_book2002}.

A \textit{Darboux transformation} for an arbitrary linear partial dif\/ferential operator $\o{L}$
can be def\/ined as follows.
An operator $\o{L}$ is transformed into operator $\o{L}_1$ with the same principal symbol
(see Section~\ref{sec:pre}) by means of operator $\o{M}$ if there is a~linear partial dif\/ferential
operator $\o{N}$ such that
\begin{gather}\label{main}
\o{N}\o{L}=\o{L}_1\o{M}.
\end{gather}
In this case we shall say that there is a~\textit{Darboux transformation for pair} $(\o{L},\o{M})$;
we also say that $\o{L}$ \textit{admits a~Darboux transformation generated by} $\o{M}$.
We def\/ine \textit{the order of a~Darboux transformation} as the order of the $\o{M}$ corresponding
to it.

If a~linear partial dif\/ferential operator $\o{L}\in K[D_x,D_y]$ has a~right factor $D_x+l$, then
it admits a
Darboux transformation generated by $\o{M}=D_x+l$.
Indeed, suppose $\o{L}=\o{F}\left(D_x+l\right)$, then for arbitrary $t\in K$:
$(D_x+t)\o{L}=(D_x+t)\o{F}(D_x+l)$, that is $\o{L}$ transforms into $(D_x+t)\o{F}$.
However, a~general Darboux transformation is not implied by a~factorization.

For operator~\eqref{op:L2XX} the only possible Darboux transformations are those
generated by $\o{M}=D_x-\psi_{1,x}\psi_1^{-1}$, where $\psi_1$ is in the kernel of the
operator~\eqref{op:L2XX}.
For operator~\eqref{op:L2} Darboux transformations can be
generated by each of the following:
\begin{align}
& \o{M}_y=D_x - \psi_{1,x} \psi_1^{-1}, \label{My} \\
& \o{M}_x=D_y - \psi_{1,y} \psi_1^{-1}, \label{Mx}
\end{align}
where $\psi_1$ is in the kernel of operator~\eqref{op:L2}.
For operator~\eqref{op:L2}, however, there
are also Darboux transformations that are generated by $\o{M}$ of a~form $D_x+m$ or $D_y+
m$, $m\in K$, but cannot be
represented as either~\eqref{My} or~\eqref{Mx}.
These transformations are known as \textit{Laplace transformations}.
Specif\/ically, these transformations are transformations of operators
$\o{L}$ of the form~\eqref{op:L2} and are generated by $\o{M}=D_x+b$, or $\o{M}=D_y+a$,
provided
the Laplace invariant $k=b_y+ab-c$ is not zero (or the invariant $h=a_x+ab-c$ is not
zero).
See more about Laplace transformations in~\cite{ts:steklov_etc:00}.
It is also known that Laplace transformations are invertible~\cite{Ganzha12}.

Moreover, it has been proved~\cite{Laplace_only_degenerate_2012} that Laplace transformations
are the only two Darboux transformations of~\eqref{op:L2} generated by $\o{M}$ of a~form $D_x+m$
or $D_y+m$, $m\in K$
that cannot be represented as either~\eqref{My} or~\eqref{Mx}.
Operators~\eqref{My} and~\eqref{Mx} can be equivalently def\/ined by their action on functions in
terms of Wronskians:
\begin{align*}
& \o{M}_x (\psi)=
\left(D_x - \frac{\psi_{1,x}}{\psi_1} \right)(\psi) =
\frac{\left|
\begin{matrix}
\psi & \partial_x\psi \\
\psi_1 & \partial_x\psi_1
\end{matrix}
\right|}{-\left| \psi_1\right|}, \\
&\o{M}_y (\psi)= \left( D_y - \frac{\psi_{1,y}}{\psi_1} \right) (\psi) =
\frac{\left|
\begin{matrix}
\psi & \partial_y\psi \\
\psi_1 & \partial_y\psi_1
\end{matrix}
\right|}{-\left| \psi_1\right|}.
\end{align*}
The consecutive application of a~sequence of $n$ Darboux transformations leads to one Darboux
transformation of order $n$,
which then can be def\/ined in terms of Wronskians of order $n$.
Indeed, if we def\/ine a~$(t,s)$ Wronkian, $t+s=n-1$, of functions $f_1,\dots,f_{t+s}$ as
the $n\times n$ determinant
\begin{gather*}
W_{t,s}(f,f_1,\dots,f_{t+s})=
\left|
\begin{matrix}
f&\partial_x f&\dots&\partial^t_x f&\partial_y f&\dots&\partial^s_y f\\
\dots&\dots&\dots&\dots&\dots&\dots&\dots\\
f_{t+s}&\partial_x f_{t+s}&\dots&\partial^t_x f_{t+s}&\partial_y
f_{t+s}&\dots&\partial^s_y f_{t+s}
\end{matrix}
\right|,
\end{gather*}
then we can reformulate Darboux's original result~\cite{Darboux2} and join it to the
result~\cite{2ndorderparab} for operators
of the form
\begin{gather}\label{op:Lxxg}
D_x D_x+a D_x+bD_y+c,\qquad a,b,c\in K
\end{gather}
as follows.
\begin{theorem} \label{DTS}
If $m+n$ linearly independent $\psi_1,\dots,\psi_{m+n}$ in the kernel of $\o{L}$ of the
form~\eqref{op:L2} or~\eqref{op:Lxxg} are given,
then operator $\o{M}$ given by
\begin{gather*}
\o{M}(\psi)
=(-1)^{n+m}
\frac{W_{m,n}(\psi,\psi_1,\dots,\psi_{m+n})}{W_{m-1,n}(\psi_1,\dots,\psi_{m+n})}
\end{gather*}
defines some Darboux transformation for $\o{L}$.
\end{theorem}

In~\cite{clouds:bagrov:samsonov:97}, which preceded~\cite{2ndorderparab}, this theorem was proved
for operators of the form~\eqref{op:Lxxg} with constant coef\/f\/icients.
The general question whether a~Darboux transformation of order $n$ can be represented as a~sequence
of Darboux transformations of order one is
open for operators of the form~\eqref{op:L2}.
In~\cite{shem:darboux2} the statement was proved
for operators of the form~\eqref{op:L2} with arbitrary (i.e.\
not necessarily constant) coef\/f\/icients, but for transformations of order two only.
For operators of the form~\eqref{op:L2XX}, the statement is an implication of the Crum
theorem~\cite{matveev:1991:darboux},
which states that a~Darboux transformation of order $n$ is generated by $\o{M}$ which can be
def\/ined by the same
Wronskian formulae (with the only dif\/ference that this is a~case of a~single variable):
\begin{gather*}
\o{M}\psi=\frac{W_{n-1,0}(\psi_1,\dots,\psi_n,\psi)}{W_{n-2,0}(\psi_1,\dots,\psi_n)}.
\end{gather*}

Generalizations of Darboux transformations have been introduced for other types of operators and
for systems of operators.
Thus, in~\cite{nimmo2010} one is def\/ined in terms of a~twisted derivation $D$ satisfying $D(AB)=
D(A)+\sigma(A)B$, where $\sigma$ is a~homomorphism.
Such twisted derivations include regular derivations, dif\/ference and
q-dif\/ference operators and superderivatives as special cases.
The corresponding $\o{M}$ can be expressed by the same formulae
in terms of Wronskians.

However, a~straightforward generalization of Theorem~\ref{DTS} would not be true for many other
kinds of operators.
This fact has not been much stressed in relevant papers.
Below is possibly the f\/irst explicit example
of such a~situation.
\begin{example} \label{ex:no_wr}
Consider operator $\o{L}=D_x^2D_y+y D_x^2+x D_y^2+1$ and an element of its kernel,
\begin{gather*}
\psi_1=\sin\left(\frac{y}{\sqrt{x}}\right)\in\ker\o{L}.
\end{gather*}
Straightforward computations show that no $\o{M}$ in the form~\eqref{My},
nor in the form~\eqref{Mx},
generates a~Darboux transformation for operator $\o{L}$.
There are, however,
other $\psi_1\in\ker\o{L}$, which generate Darboux transformations with
$\o{M}$ in the form~\eqref{My} or~\eqref{Mx}.
\end{example}

In the present paper we are interested in answering the following questions.
\begin{enumerate}\itemsep=0pt
\item Laplace transformations are known to be invertible, but essentially no other invertible
transformations are known.
Are there such transformations? Can we study their nature?
\item Example~\ref{ex:no_wr} indicates that Wronkian formulae do not work in some cases.
Can we describe cases
in which Wronkian formulae do work?
\end{enumerate}

Specif\/ically, in the present paper we show that there are many more kinds of operators than just
the one~\eqref{op:L2}
which admit invertible (see the precise def\/inition in Section~\ref{sec:inv}) Darboux
transformations.
We give criteria which allow us to describe all possible invertible Darboux transformations of
arbitrary linear partial dif\/ferential operators
$\o{L}$ and generated by $\o{M}$ in the form $D_x+m$ or $D_y+m$.
Given an arbitrary operator $\o{L}\in K[D_x,D_y]$, we give suf\/f\/icient conditions that guarantee
that Wronskian
formulae~\eqref{My} and~\eqref{Mx} work.

This paper is organized as follows.
Section~\ref{sec:pre} outlines notation.
Section~\ref{sec:dxdy} starts with the simple fact that a~Darboux transformation for a
pair $(\o{L},\o{M}=D_x+m)$, $m\in K$ exists if and only if a~Darboux transformation for
pair $(\o{L}^g,D_x)$ exists (analogously, for
pairs with $\o{M}=D_y+m$).
Then Theorem~\ref{thm:dx:dy} states necessary and suf\/f\/icient conditions that guarantee
that a~Darboux transformation for pair $(\o{L}^g,D_x)$ exists.
In Section~\ref{sec:inv} we def\/ine invertible transformations as transformations such
that the corresponding mapping $\ker\o{L}\to\ker\o{L}_1$ is invertible.
Theorem~\ref{thm:dim} states necessary and suf\/f\/icient conditions that guarantee that a~Darboux
transformation for pair $(\o{L}^g,D_x)$ is invertible
(as well as for pairs with $\o{M}=D_y+m$).
It also describes all other possible cases for the dimension of the kernel
of mapping $\ker\o{L}\to\ker\o{L}_1$.
In Section~\ref{sec:wr} Theorem~\ref{thm:mystery} states that if in the kernel of operator
$\o{L}$ there is a~subspace of large enough dimension generated by elements that dif\/fer from each
other by a~multiplication of
a function of variable $y$, then each of those elements $\psi_1$ gives the same
$\o{M}=D_x-\psi_{1,x}\psi_1^{-1}$, and this $\o{M}$
generates a~Darboux transformation.
In other words, we state when Darboux Wronskian formulae work for arbitrary bivariate linear partial
dif\/ferential operator.
There is also an analogous statement for the case $\o{M}=D_y-\psi_{1,y}\psi_1^{-1}$.

\section{Preliminaries}\label{sec:pre}

Let $K$ be a~dif\/ferential f\/ield of characteristic zero with commuting
derivations $\partial_x$, $\partial_y$.
Let $K[D_x,D_y]$ be the corresponding ring of linear
partial dif\/ferential operators over~$K$, where~$D_x$,~$D_y$ correspond to
derivations~$\partial_x$,~$\partial_y$.

Operators $\o{L}\in K[D_x,D_y]$ have the general form
\begin{gather}\label{op:Ln}
\o{L}=\sum_{i+j=0}^d a_{ij}D_x^i D_y^j,\qquad a_{ij}\in K.
\end{gather}
The formal polynomial
\begin{gather*}
\Sym_\o{L}=\sum_{i+j=d}a_{ij}X^i Y^j
\end{gather*}
in the formal variables $X$, $Y$ is called the \emph{principal symbol} of $\o{L}$.

One can either assume the f\/ield $K$ to be dif\/ferentially closed, in other words containing all the
solutions
of, in general nonlinear, partial dif\/ferential equations
with coef\/f\/icients in~$K$, or simply assume that~$K$ contains the solutions of those partial
dif\/ferential equations that we encounter on the way.

Let $f\in K$, and $\o{L}\in K[D_x,D_y]$; by $\o{L}f$ we denote the composition of
operator $\o{L}$ with the operator of multiplication
by a~function $f$, while $\o{L}(f)$ mean the application of operator~$\o{L}$ to~$f$.
For example,
\begin{gather*}
D_x f = f D_x + f_x \in K[D_x, D_y],\qquad
D_x (f) = f_x \in K.
\end{gather*}
The second lower index attached to a~symbol denoting a~function means the derivative of that
function with respect to the variables listed there.
For example,
\begin{gather*}
f_{1,xyy}=\partial_x\partial_x\partial_y f_1.
\end{gather*}

\begin{definition} Given some operator $\o{R}\in K[D_x,D_y]$ and an invertible function $g\in
K$, the corresponding
\textit{gauge transformation} is def\/ined as
\begin{gather*}
\o{R}\to\o{R}^{g},\qquad \o{R}^g=g^{-1}\circ R\circ g,
\end{gather*}
where $\circ$ denotes the operation of the composition of operators in $K[D_x,D_y]$.
\end{definition}
\begin{remark}
The principal symbol of an operator in $K[D_x,D_y]$ is invariant under the gauge transformations.
\end{remark}

\begin{lemma}
\label{lem:conj:preserves:DT}
Let $\o{M},\o{L}\in K[D_x,D_y]$ and let a~Darboux transformation exist for the
pair $(\o{M},\o{L})$.
Then one exists also for $(\o{M}^g,\o{L}^g)$, where $g$ is an arbitrary invertible element of $K$.
\end{lemma}
\begin{proof} Indeed,~\eqref{main} implies
$g^{-1}\circ\o{N}\circ g\circ g^{-1}\circ\o{L}\circ g$ $=g^{-1}\circ\o{L}_1\circ g\circ
g^{-1}\circ\o{M}\circ g$,
and, therefore, $\o{N}^g\circ\o{L}^g=\o{L}_1^g\circ\o{M}^g$.
Since gauge transformations do not change the symbol of an operator,
the proof is complete.
\end{proof}

\section[Darboux transformations generated by $\o{M}=D_x$ or $\o{M}=D_y$]{Darboux transformations generated by $\boldsymbol{\o{M}=D_x}$ or $\boldsymbol{\o{M}=D_y}$}\label{sec:dxdy}

\begin{lemma}
Let $\o{L}$ be an arbitrary linear partial differential operator in $K[D_x,D_y]$.
If there exists a~Darboux transformation for the pair $(\o{L},\o{M}=D_x+m)$, $m\in K$
then there is a~Darboux transformation for pair $(\o{L}^g,D_x)$.
If there exists a~Darboux transformation for pair $(\o{L},\o{M}=D_y+m)$,
then there exists a~Darboux transformation for pair $(\o{L}^g,D_y)$.
\end{lemma}
\begin{proof}
Consider a~Darboux transformation for the pair $(\o{L},\o{M}=D_x+m)$,
and consider $g$, a~solution of $g_x/g=-m$.
Then $\o{M}^g=D_x$.
Then by Lemma~\ref{lem:conj:preserves:DT} there is also a~Darboux transformation for
pair $(\o{L}^g,D_x)$.
Analogous reasoning applies for the pair $(\o{L},\o{M}=D_y+m)$.
\end{proof}
\begin{theorem}\label{thm:dx:dy}
Let $\o{L}$ be an arbitrary linear partial differential operator in $K[D_x,D_y]$, that is, an
operator of the form~\eqref{op:Ln}.

Operator $\o{L}$ has a~Darboux transformation generated by $\o{M}=D_x$ if and only if
\begin{gather}\label{cond1n}
a_{0j}=G_{jk}(y)a_{0k}
\end{gather}
for all non-zero $a_{0j}$ and $a_{0k}$, $j=0,\dots,d$, $k=0,\dots,d$.
Here $G_{jk}(y)$
are some functions of the variable $y$.

Operator $\o{L}$ has a~Darboux transformation generated by $\o{M}=D_y$ if and only if
\begin{gather}\label{cond2n}
a_{i0}=F_{ik}(x)a_{k0}
\end{gather}
for all non-zero $a_{i0}$ and $a_{k0}$, $i=0,\dots,d$, $k=0,\dots,d$.
Here $F_{ik}(x)$
are some functions of the variable $x$.
\end{theorem}
\begin{proof}
Let there be a~Darboux transformation for the pair $(\o{L},D_x)$.
This means that for some $\o{L}_1\in K[D_x,D_y]$ and $n\in K$
\begin{gather}\label{Lndt}
(D_x+n)\o{L}=\o{L}_1D_x.
\end{gather}
The right hand side of this operator equality does not contain terms of the form $c D_y^j$, $c\in
K$.
On the other hand for $j=1,\dots,d$ the coef\/f\/icient of $D_y^j$ in the left hand side is
\begin{gather}\label{a0j}
n a_{0j}+a_{0j,x}=0.
\end{gather}
The `free' coef\/f\/icient on the right hand side of~\eqref{Lndt} is zero, while on the left hand side
it is
\begin{gather}\label{a00}
n a_{00}+a_{00,x}=0.
\end{gather}
Equating the expressions for $n$ obtained from each of the equalities~\eqref{a0j} and~\eqref{a00},
we see that condition~\eqref{cond1n} is satisf\/ied.

The analogous statement for the case $\o{M}=D_y$ can be proved similarly.

To prove the theorem in the other direction, suppose that the conditions~\eqref{cond1n} are
satisf\/ied.
Consider then the operator
equality~\eqref{Lndt} def\/ining a~Darboux transformation for the pair $(\o{L},D_x)$.
It implies equalities~\eqref{a0j},~\eqref{a00}, from which
$n$ can be determined uniquely and without contradiction.
In general, equality~\eqref{Lndt}, which is an equality of operators of orders $d+1$, implies
$(d+3)(d+2)/2$ equalities of the corresponding coef\/f\/icients, of which $(d+1)$ are
equalities~\eqref{a0j} and~\eqref{a00}.
Thus,
we have a~linear algebraic system of a~maximum
$(d+3)(d+2)/2-(d+1)$ equations to solve.
The unknowns in this system are the coef\/f\/icients of $\o{L}_1$,
an operator of order~$d$ and their number is~$(d+2)(d+1)/2$.
Thus the number of equations in this system is less than or equal to the number of unknowns,
so there is at least one non-zero solution.

Analogously we can prove that conditions~\eqref{cond2n} guarantee that there is a~Darboux
transformation for pair $(\o{L},D_y)$.
\end{proof}

\section{Invertible Darboux transformations}\label{sec:inv}

Consider a~Darboux transformation $\o{L}\to\o{L}_1$ of operators in $K[D_x,D_y]$ generated by
an operator $\o{M}\in K[D_x,D_y]$.
This transformation implies a~mapping of the linear vector spaces
\begin{gather*}
\ker\o{L}\to\ker\o{L}_1:\ \  w\mapsto\o{M}(w).
\end{gather*}
This mapping is invertible if it is an isomorphism, that is if its kernel is zero
\begin{gather*}
\ker\o{L}\cap\ker\o{M}=\{0\}.
\end{gather*}
In this case we shall say that the \textit{Darboux transformation is invertible}.

\begin{remark} Darboux transformations of the operator $\o{L}\in K[D_x,D_y]$ generated by $\o{M}$
constructed using Wronkian
formulae are not invertible.
Indeed, we take
some number of known solutions of $\o{L}(\psi)=0$ and then construct $\o{M}$ as a~Wronkian.
Then
by construction all those solutions belong to the kernel of $\o{M}$.
\end{remark}

\begin{theorem}\label{thm:dim}
If the operator $\o{L}\in K[D_x,D_y]$ given by~\eqref{op:Ln} admits a~Darboux transformation
generated by $\o{M}=D_x$ or $\o{M}=D_y$ then
\begin{gather*}
\dim(\ker(\o{L})\cap\ker(\o{M}))=
\begin{cases}
\infty,&a_{0k}=0,\quad\forall\, k\in\{0,\dots,d\},\\
d_y,& \text{otherwise},
\end{cases}
\end{gather*}
where $d_y$ is the largest $j$ such that $a_{0j}\neq0$.
In particular, we have  a criteria for a~Darboux transformation  generated by $\o{M}=D_x$
or $\o{M}=D_y$
to be invertible.
Namely, it is invertible if and only if all of the following hold
\begin{gather*}
a_{0k}=0\quad\forall\,  k\in\{1,\dots,d\},\qquad
a_{00}\neq0.
\end{gather*}
\end{theorem}
\begin{proof}
Suppose there is a~Darboux transformation for the pair $(\o{L},D_x)$.
The kernel of the operator $\o{M}$ consists of functions of the form $g=g(y)$.

Let there be index $k$, $k>0$ such that the coef\/f\/icients of $D_y^k$, $k=1,\dots,d$ are not
zero: $a_{0k}\neq0$.
Then using~\eqref{cond1n} compute
\begin{gather*}
\o{L}(g(y))=\sum_{j=0}^d a_{0j}D^j_y(g(y))
=a_{0k}\sum_{j=0}^d G_{jk}(y)D^j_y(g(y)).
\end{gather*}
Thus, $\o{L}(g(y))=0$ is equivalent to a~linear dif\/ferential equation of order $d_y$.
The solution space of such an equation has dimension $d_y$.

If there is no such $k$ that the coef\/f\/icient at $D_y^k$, $k=1,\dots,d$ is not zero, then
either $a_{00}=0$ and then $\ker\o{L}\cap\ker\o{M}=\{f(y)\}$, or
$a_{00}\neq0$ and $\ker\o{L}\cap\ker\o{M}=\{0\}$.

The statement for the pair~$(\o{L},D_y)$ can be proved analogously.
\end{proof}

\begin{corollary}
Laplace transformations $($see Section~{\rm \ref{sec:intr})} are the only invertible Darboux transformation
generated by $\o{M}=D_x+m$
or by $\o{M}=D_y+m$ for operators in the form $D_xD_y+a D_x+bD_y+c$, $a,b,c\in K$.
\end{corollary}

\begin{proof} In~\cite{Laplace_only_degenerate_2012} it has been proved that a~Darboux
transformation
generated by $\o{M}=D_x+m$ or by $\o{M}=D_y+m$ is either a~Laplace transformation, that is
generated by $\o{M}=D_x+b$ or $\o{M}=D_y+a$,
or generated by $\o{M}$ in the form $D_x-\psi_{1,x}\psi_1^{-1}$,
or $D_y-\psi_{1,y}\psi_1^{-1}$, where $\psi_1\in\ker\o{L}$.
In the latter case, $\psi_1\in\left(\ker\o{L}\cap\ker\o{M}\right)$, and, therefore, the
mapping
$\ker\o{L}\to\ker\o{L}_1$ is not invertible.

Consider a~Darboux transformation for pair $(\o{L},\o{M}=D_x+b)$.
To use the criteria
obtained in Theorem~\ref{thm:dim}, we consider a~gauge transformation of both operators of this
pair.
We use $g\in K$ such that $\o{M}$ becomes $D_x$, that is $\o{M}^g=D_x$.
Such $g$ can be easily found as any solution of $g_x g^{-1}=-b$.
Consider now $\o{L}^g$,
\begin{gather*}
\o{L}^g  = D_xD_y + \big(a+g_y g^{-1}\big) D_x + \big(b+g_x g^{-1}\big) D_y + c + a~g_x g^{-1} + b g_y g^{-1} +
g_{xy} g^{-1}  \\
\hphantom{\o{L}^g }{} = D_xD_y + \big(a+g_y g^{-1}\big) D_x + c - a~b + b g_y g^{-1} + g_{xy} g^{-1}  \\
\hphantom{\o{L}^g }{}= D_xD_y + \big(a+g_y g^{-1}\big) D_x + c - a~b - b_y  \\
\hphantom{\o{L}^g }{}= D_xD_y + \big(a+g_y g^{-1}\big) D_x + k,
\end{gather*}
where $k$ is one of two Laplace invariants of $\o{L}$ (see Section~\ref{sec:intr}).
Thus, $k\neq0$ is the necessary and suf\/f\/icient condition
for the Darboux transformation for the pair $(\o{L},\o{M}=D_x+b)$ to be invertible.
Analogously, $h\neq0$ is the necessary and suf\/f\/icient condition for Darboux transformation for
pair $(\o{L},\o{M}=D_y+a)$ to be invertible.
\end{proof}

\begin{example}[an invertible Darboux transformation]
Consider the operator
\begin{gather*}
\o{L}=D_x D_y^2+D_x^2+x D_x+1.
\end{gather*}
The coef\/f\/icients of $\o{L}$ satisfy condition~\eqref{cond1n}, and, therefore, there is
a Darboux transformation for $\o{L}$ generated by $\o{M}=D_x$.
Since there is no term in the form $D_y^i$ and $\o{L}(1)=1\neq0$, then by Theorem~\ref{thm:dim}
the Darboux transformation for pair $(\o{L},\o{M})$ is invertible.
This Darboux transformation takes $\o{L}$ into
\begin{gather*}
\o{L}_1=D_x D_y^2+D_x^2+x D_x+2.
\end{gather*}
The corresponding operator $\o{N}$ is $D_x$.
In this case there is an inverse transformation of operators, $\o{L}_1\mapsto\o{L}$ by means
of $\o{M}'=\o{N}'=D_x$.

From any such example one can generate a~whole series of examples of Darboux transformations
with $\o{M}$
of the form $D_x+m$ or $D_y+m$, $m\neq0$ by considering gauge transformations of $\o{L}$
and $\o{M}$.
\end{example}

\section{Solution-based Darboux transformations}\label{sec:wr}

\begin{theorem}\label{thm:mystery}
Let an operator $\o{L}$ be of the form $\o{L}=\sum\limits_{i+j=0}^d a_{ij}D_x^i D_y^j$, $a_{ij}\in
K$,
and $k=d$ if $a_{0d}\neq0$, and $k=d-1$, otherwise.
Then if there are
\begin{gather*}
\psi_1,\psi_2,\dots,\psi_{k}\in\ker\o{L}/\{0\}
\end{gather*}
such that
\begin{gather}\label{eq:cond:sols}
\frac{\psi_i}{\psi_1}=T_i(y),
\end{gather}
for some non-constant functions $T_i(y)\in K$, $i=2,\dots,k$, and
\begin{gather}\label{cond:wr_not_0}
W_{0,k-1} (1,T_2,\dots,T_{k} )\neq0,
\end{gather}
then there is a~Darboux transformation for $\o{L}$ generated by
\begin{gather*}
\o{M}=D_x-\frac{\psi_{1,x}}{\psi_1}.
\end{gather*}
Analogously, if instead of~\eqref{eq:cond:sols} and~\eqref{cond:wr_not_0}
conditions $\psi_i\psi_1^{-1}=F_i(x)$
and $W_{k-1,0} (1,T_2,\dots,T_{k} )\neq0$ hold, then there is a~Darboux
transformation for $\o{L}$ generated by
$\o{M}=D_y-\psi_{1,y}\psi_1^{-1}$.
\end{theorem}

\begin{remark}
Condition~\eqref{cond:wr_not_0} implies that $\psi_1,\psi_2,\dots,\psi_{k}$ are linearly
independent.
The condition that
$\psi_1,\psi_2,\dots,\psi_{k}$ are linearly independent
does not necessarily imply condition~\eqref{cond:wr_not_0}.
\end{remark}
\begin{proof}
Assume we have non-zero $\psi_1,\psi_2,\dots,\psi_{k}\in\ker\o{L}$, such
that~\eqref{eq:cond:sols} and~\eqref{cond:wr_not_0} hold.
Then for the operator $\o{L}'=\o{L}^{\psi_1}=\sum_{i+j=0}^d a'_{ij}D_x^i D_y^j$, $a'_{ij}\in K$
we have
\begin{gather*}
1\in\ker\o{L}',\qquad T_i(y)\in\ker\o{L}',\qquad i=1,\dots,k.
\end{gather*}
The former means that $a'_{00}=0$, while the latter implies that
\begin{gather}\label{sys:gy}
\sum_{j=1}^{d}a'_{0j}D^j_y(T_i(y))=0,\qquad i=2,\dots,k\
\end{gather}
which is a~linear system of $k-1$ equations.
The number of unknowns $a'_{0j}$ is the number of
non-zero $a'_{0j}$, $j=1,\dots,d$.
If $k=d$, then the number of the unknowns is less or equal to $k$.
If $k=d-1$, then $a_{0d}=0$, and since the principal symbol of an operator is invariant
with respect to the gauge transformations, $a'_{0d}=a_{0d}=0$.
Which means that in this case
the number of unknowns is also less than or equal to $k$.

The Wronkian in~\eqref{cond:wr_not_0} has the f\/irst column $(1,0,\dots,0)^t$, and the f\/irst
row $(1,T_2,\dots T_{k})$.
Thus, if we remove the f\/irst column and the f\/irst row, the rank of the corresponding matrix
is $k-1$.
Then the rank of the transpose of the latter matrix
is also $k-1$,
\begin{gather*}
\rk\left(
\begin{matrix}
T_2'&\dots&T_2^{(k-1)}\\
\dots&\dots&\dots\\
T_{k}'&\dots&T_{k}^{(k-1)}
\end{matrix}
\right)=k-1.
\end{gather*}
This is a~$(k-1)\times(k-1)$ matrix and adding extra column $(T_2^{k},\dots,T_{k}^{k})^t$ on the
right does not change the rank.
Thus linear system~\eqref{sys:gy} has matrix with rank $k-1$ and less or equal to $k$ unknowns,
and, therefore,
condition
\begin{gather*}
a'_{0j}=G_{jk}(y)a'_{0k}
\end{gather*}
holds
for all non-zero $a'_{0j}$ and $a'_{0k}$, $j=0,\dots,d$, $k=0,\dots,d$ for some $G_{jk}(y)\in
K$.
Thus, by Theorem~\ref{thm:dx:dy}
there exists a~Darboux transformation for the pair $(\o{L}',D_x)$.
Therefore, there is
a~Darboux transformation for pair $(\o{L},\psi_1D_x\psi_1^{-1})$, that is for pair $(\o{L},
D_x-\psi_{1,x}\psi_1^{-1})$.

The second statement, giving suf\/f\/icient conditions for the existence of a~Darboux transformation
for $\o{L}$ generated by
$\o{M}=D_y-\psi_{1,y}\psi_1^{-1}$, can be proved analogously.
\end{proof}
\begin{example}
For operators of the form $D_xD_y+a D_x+bD_y+c$, $a,b,c\in K$
$k=1$, Theorem~\ref{thm:mystery} then means that it is enough
to have one $\psi\in\ker\o{L}$ to guarantee that a~Darboux transformation for
pair $(\o{L},\o{M}=D_x-\psi_{1,x}\psi_1^{-1})$
exists.
This agrees with result of Theorem~\ref{DTS}.
\end{example}
\begin{remark}
The statement of Theorem~\ref{thm:mystery} is not true when formulated in the opposite direction.
Let there be a~Darboux transformation for the pair $(\o{L},\o{M}=D_x-\psi_{1,x}\psi_1^{-1})$,
where $\psi_1\in\ker\o{L}$.
Then there is a~Darboux transformation for the
pair $(\o{L}'=\o{L}^{\psi_1},\o{M}'=\o{M}^{\psi_1}=D_x)$,
$a'_{00}=0$.
If~$d'_y$ is the largest $j$ such that $a'_{0j}\neq0$, and since $a'_{00}=0$
Theorem~\ref{thm:dim} implies that there are either~$d'_y$ (or inf\/initely many) linearly independent
\begin{gather}\label{seq:prime}
u_i\in\ker\o{L}'\cap\ker\o{M}',\qquad i=1,\dots, d'_y\qquad(\text{or } i=1,\dots,\infty).
\end{gather}
Since $\o{M}'=D_x$, each of them is a~function of the variable $y$ solely.
Since $1\in\ker\o{L}\cap\ker\o{L}$, we can choose~\eqref{seq:prime}
in such a~way that $u_1=1$.
Then
\begin{gather*}
\phi_1=\psi_1,\, \phi_{2}=\psi_1u_{2},\,\ldots\in\ker\o{L}\cap\ker\o{M},
\end{gather*}
and $\psi_i/\psi_1$, $i=1,\dots,d'_y$ are some function of $y$ solely.
Thus, if the number $d'_y$ is less then $k$, we cannot get $k$
\begin{gather*}
\psi_1,\psi_2,\dots,\psi_{k}\in\ker\o{L}\cap\ker\o{M}.
\end{gather*}
\end{remark}

\begin{example}[the largest $j$ such that $a_{0j}\neq0$ is not invariant under gauge
transformation]
Consider the operator
\begin{gather*}
\o{L}=D_x^2D_y+D_y^2D_x-\frac{1}{x}D_y^2+D_y,
\end{gather*}
and $\psi_1=x\in\ker\o{L}$.
Then
\begin{gather*}
\o{L}'=\o{L}^{\psi_1}=D_x^2D_y+D_y^2D_x+\frac{2}{x}D_x D_y+D_y.
\end{gather*}
Thus, $d_y=2$ and $d_y'=1$, and there exists (a non-invertible)
Darboux transformation generated by $\o{M}=D_x-\psi_{1,x}\psi_1^{-1}=D_x-\frac{1}{x}$.
It takes $\o{L}$ into
\begin{gather*}
\o{L}_1=D_x^2D_y+D_y^2D_x-\frac{1}{x}D_y^2+\left(1-\frac{2}{x^2}\right)D_y,
\end{gather*}
and the corresponding $\o{N}$ is $\o{N}=\o{M}$.
\end{example}

Theorem~\ref{thm:mystery} states some conditions under which an operator has non-invertible
Darboux transformations generated by $\o{M}$ in the form $D_x-\psi_{1,x}\psi_1^{-1}$
or $D_y-\psi_{1,y}\psi_1^{-1}$,
where $\psi_1\in\ker\o{L}$.

\section{Conclusions}

The paper provides the f\/irst ideas for the general (algebraic) theory of
invertible f\/irst-order Darboux transformations for bivariate linear partial dif\/ferential operators
of arbitrary order $d$
and of arbitrary form.
Although the order of the auxiliary operator has been restricted to the f\/irst order, the major
operator is taken in general form.

Future work may include strengthening of Theorem~\ref{thm:mystery} to describe a~criteria.
It is not known yet how to extend these ideas to the important three dimensional case,
the generalized Laplace transformations for which have been recently developed in~\cite{Ganzha12}.
It is also extremely important to extend these ideas for discrete analogies that are under active
development
now~\cite{grinevich_novikov_2012discrete, novikov_in_mikhailov2008integrability}.

\pdfbookmark[1]{References}{ref}
\LastPageEnding


\begin{thebibliography}{99}
\footnotesize\itemsep=0pt

\bibitem{clouds:bagrov:samsonov:97}
Bagrov V.G., Samsonov B.F., Darboux transformation of the {S}chr\"odinger
  equation, \href{http://dx.doi.org/10.1134/1.953045}{\textit{Phys. Part. Nuclei}} \textbf{28} (1997), 374--397.

\bibitem{Darboux2}
Darboux G., Le\c{c}ons sur la th{\'e}orie g{\'e}n{\'e}rale des surfaces et les
  applications g{\'e}om{\'e}triques du calcul inf\/init{\'e}simal.~II, Gauthier-Villars, Paris, 1889.

\bibitem{Ganzha12}
Ganzha E.I., On Laplace and Dini transformations for multidimensional equations
  with a decomposable principal symbol, \href{http://dx.doi.org/10.1134/S0361768812030012}{\textit{Program. Comput. Softw.}}
  \textbf{38} (2012), 150--155.

\bibitem{grinevich_novikov_2012discrete}
Grinevich P.G., Novikov S.P., Discrete $SL_2$ connections and self-adjoint
  dif\/ference operators on the triangulated 2-manifold, \href{http://arxiv.org/abs/1207.1729}{arXiv:1207.1729}.

\bibitem{nimmo2010}
Li C.X., Nimmo J.J.C., Darboux transformations for a twisted derivation and
  quasideterminant solutions to the super {K}d{V} equation, \href{http://dx.doi.org/10.1098/rspa.2009.0647}{\textit{Proc.~R.
  Soc. Lond. Ser.~A Math. Phys. Eng. Sci.}} \textbf{466} (2010), 2471--2493,
  \href{http://arxiv.org/abs/0911.1413}{arXiv:0911.1413}.

\bibitem{matveev:1991:darboux}
Matveev V.B., Salle M.A., Darboux transformations and solitons, \textit{Springer Series
  in Nonlinear Dynamics}, Springer-Verlag, Berlin, 1991.

\bibitem{novikov_in_mikhailov2008integrability}
Novikov S.P., Four lectures: discretization and integrability. Discrete
  spectral symmetries, in Integrability, \href{http://dx.doi.org/10.1007/978-3-540-88111-7_4}{\textit{Lecture Notes in Physics}},
  Vol.~767, Editor A.V.~Mikhailov, Springer, Berlin, 2009, 119--138.

\bibitem{novikov1997exactly}
Novikov S.P., Veselov A.P., Exactly solvable two-dimensional {S}chr\"odinger
  operators and {L}aplace transformations, in Solitons, geometry, and topology:
  on the crossroad, \textit{Amer. Math. Soc. Transl. Ser.~2}, Vol.~179, Amer.
  Math. Soc., Providence, RI, 1997, 109--132, \href{http://arxiv.org/abs/math-ph/0003008}{math-ph/0003008}.

\bibitem{rogers_schief_book2002}
Rogers C., Schief W.K., B\"acklund and {D}arboux transformations. Geometry and
  modern applications in soliton theory, \href{http://dx.doi.org/10.1017/CBO9780511606359}{\textit{Cambridge Texts in Applied
  Mathematics}}, Cambridge University Press, Cambridge, 2002.

\bibitem{Laplace_only_degenerate_2012}
Shemyakova E., Laplace transformations as the only degenerate {D}arboux
  transformations of f\/irst order, \href{http://dx.doi.org/10.1134/S0361768812020053}{\textit{Program. Comput. Softw.}} \textbf{38}
  (2012), 105--108.

\bibitem{shem:darboux2}
Shemyakova E., Proof of the completeness of {D}arboux {W}ronskian formulas for
  order two, \href{http://dx.doi.org/10.4153/CJM-2012-026-7}{\textit{Canad.~J. Math.}}, {t}o appear, \href{http://arxiv.org/abs/1111.1338}{arXiv:1111.1338}.


\bibitem{ts:steklov_etc:00}
Tsarev S.P., Factorization of linear partial dif\/ferential operators and the
  {D}arboux method for integrating nonlinear partial dif\/ferential equations,
  \href{http://dx.doi.org/10.1007/BF02551175}{\textit{Theoret. Math. Phys.}} \textbf{122} (2000), 121--133.

\bibitem{2ndorderparab}
Tsarev S.P., Shemyakova E., Dif\/ferential transformations of second-order
  parabolic operators in the plane, \href{http://dx.doi.org/10.1134/S0081543809030134}{\textit{Proc. Steklov Inst. Math.}}
  \textbf{266} (2009), 219--227, \href{http://arxiv.org/abs/0811.1492}{arXiv:0811.1492}.

\end{thebibliography}
\end{document}